\documentclass{article}
\usepackage[utf8]{inputenc}
\usepackage{amsfonts}
\usepackage{amsmath}
\usepackage{amsfonts}
\usepackage{graphicx}
\usepackage{float}

\usepackage[
    backend=biber,
    style=nature,
  ]{biblatex}
\title{A Coherent Bi-Directional Virtual Detector for the 1-D Schr\"odinger Equation}
\author{Joshua Mann, James Rosenzweig\\
University of California, Los Angeles}

\begin{document}

\maketitle

\begin{abstract}
    The virtual detector is a commonly utilized technique to measure the properties of a wavefunction in simulation. One type of virtual detector measures the probability density and current at a set position over time, permitting an instantaneous measurement of momentum at a boundary. This may be used as the boundary condition between a quantum and a classical simulation. However, as a tool for measuring spectra, it possesses several problems stemming from its incoherent nature. Another form of virtual detector measures the wavefunction's complex value at a set position in real space over time and Fourier analyzes it to produce an energy spectrum. The spectra it produces are exact provided that the wavefunction propagated through the detector in one direction. Otherwise it will produce a spectrum that includes interference between forward and backward propagating wavepackets. Here we propose a virtual detector which maintains all the benefits of this coherent virtual detector while also being able to resolve the direction of propagation and mitigate nonphysical interference by use of a second measurement point. We show that, in the continuum limit, this bi-directional virtual detector can reproduce an equivalent wavefunction assuming a globally constant potential. It is therefore equivalent to the exact spectrum.
\end{abstract}

\section{Introduction}

In calculations utilizing the Schr\"odinger equation a common tool for measurement is the virtual detector (VD) \cite{cvdbdy, qvd1, qvd2, qvd3}. The VD permits measurement of certain properties of the wavefunction, typically at some position in real space, without perturbing the wavefunction itself. In this regard it is entirely a theoretical tool, hence why it is deemed ``virtual'' for its unperturbing measurement.

In some cases VDs take the form of a classical detector (CVD) where the wavefunction's probability density and current are measured at some boundary manifold \cite{cvdbdy}. The instantaneous classical momentum of the probability fluid may then be calculated and binned into a final spectrum. Such a result allows for the creation of a semi-classical model, with one region treated quantum mechanically and the other treated classically with phase integration \cite{cvdbdy2}. The proof \cite{cvdproof} that this equates to measuring the true momentum distribution requires that the measurement time tends towards infinity, leading to the requirement of a strong momentum-spatial correlation. An example issue here is that two opposing wavepackets of opposite central momentum that overlap at the VD will net a zero probability current and therefore a zero momentum, even though we clearly have non-zero momenta involved. To remedy this a coherent version of the VD should be considered.

One further step is to consider, in 1-D, the wavefunction's value at a single point for all time, $\psi_{x_0}(t)$ with a quantum VD (QVD) \cite{qvdog, qvd1, qvd2, qvd3}. Considering the straightforward relation $E=\hbar\omega$ one may Fourier transform the wavefunction at this point to obtain an energy spectrum at this point, $|\widetilde{\psi}_{x_0}(\omega)|^2$. Multiplying by $\sqrt{E}$, the velocity, provides the flux for each energy. However, this method has no mechanism for discerning rightward and leftward moving wavepackets; both wavepackets contribute coherently to the same part of the spectrum. Additionally, if one were to use a VD as a boundary to a classical calculation \cite{cvdbdy} the QVD would not provide an instantaneous boundary condition, as a temporal window of sufficient size is necessary to resolve the energies involved. If the user is certain that wavepackets will only travel in one direction (for instance computationally, if a good absorptive boundary is on the other side of the VD) and only expects a final spectrum then this method is sufficient for measurement.

The coherent bi-directional quantum VD (BQVD) we propose in this paper is a direct improvement upon the QVD in that it is able to coherently resolve wavepackets traveling both directions. An equivalent free-space wavefunction may then be extracted from this calculation, ultimately providing a less restrictive proof of equivalence in 1-D. The proof may be extended straightforwardly to $n$-D using a hyperplanar boundary. However, $n$-D results with more utility would include enclosed boundaries or periodic boundary conditions which may constitute future work.

\section{Derivation}

Here we derive the discrete version of the BQVD. Let us consider a standard QVD at $x=0$ with potential $V_0(t)=0$, without loss of generality, with the Fourier transform $f_0(k)=\mathcal{F}\psi(0,t)$ after directly converting from energy to a strictly positive momentum. The potential may be artificially set to zero by dividing the wavefunction by the phase accumulated at this point prior to Fourier transformation, $\exp{\left(-i/\hbar \int_0^t dt\;V_0(t)\right)}$. We represent $f_0$ as a combination of rightward $\tilde r(k)$ and leftward $\tilde l(k)$ moving waves:

\begin{equation}
    f_0(k)=\tilde r(k)+\tilde l(k)
\end{equation}

which are all functions of the absolute value of the momentum. To further discern the two let us evaluate a QVD at a nearby position $x=\Delta x$, providing $f_1(k)$,

\begin{equation}
    f_1(k)=\tilde r(k)e^{ik\Delta x} + \tilde l(k)e^{-ik\Delta x}
\end{equation}

where we assumed free-space propagation between these two virtual detectors. The two equations may then be straightforwardly solved for the directional components,

\begin{equation}
    \begin{aligned}
        \tilde r &= i\frac{\phi^-f_0-\phi^+f_1}{2\sin k\Delta x} \\
        \tilde l &= i\frac{\phi^+f_0-\phi^-f_1}{2\sin -k\Delta x}
    \end{aligned}
\end{equation}

with $\phi^\pm=e^{\pm ik\Delta x/2}$. We note that these two equations are the same aside from opposite signs of $k$ so a full spectrum may be derived by setting $f_i$ to be even,

\begin{equation}
    \tilde\psi(k)=\frac{\hbar k}{m}\tilde g(k) = i\frac{\hbar k}{2m\sin k\Delta x}(\phi^-f_0(k)-\phi^+f_1(k))
\end{equation}

which is the desired result.

\subsection{Continuous Result}

For analytical purposes this equation may be reformulated in the limit that $\Delta x\rightarrow 0$, resulting in,

\begin{equation}
    \tilde\psi(k)=\frac{\hbar}{2m}(kf(k)-i\partial_xf(k))
    \label{eqcont}
\end{equation}

with $f$ the wavefunction's Fourier transform at the VD and $\partial_xf$ the Fourier transform of the spatial derivative of the wavefunction at the VD.

\section{Proof of Equivalence}

Let us consider an arbitrary free wavefunction,

\begin{equation}
    \psi_0(x,t;\mathbf{r}_\perp)=\frac{1}{\sqrt{2\pi}}\int dk\; \tilde\psi_0(k;\mathbf{r}_\perp)e^{ikx-i\omega t}
\end{equation}

Where extra dimensions may be included within $\mathbf{r}_\perp$. We measure $\psi_0$ and $\partial_x\psi_0$ at $x=0$ for all $\mathbf{r}_\perp\in\mathbb{R}^{n-1}$. The dependence on $\mathbf{r}_\perp$ may be straightforwardly propagated through the following results to show equivalence in $n$-D, so we drop the transverse dependence for clarity. Now we may evaluate $f$ and $\partial_xf$, and apply the dispersion relation $\omega=\frac{\hbar}{2m}k^2$,

\begin{equation}
    \begin{aligned}
        f=&\frac{1}{2\pi}\int\int dk'dt\; \tilde\psi_0(k')e^{-i \frac{\hbar}{2m}(k'^2-k^2)t} \\
        \partial_xf=&\frac{i}{2\pi}\int\int dk'dt\; k'\tilde\psi_0(k')e^{-i \frac{\hbar}{2m}(k'^2-k^2)t}
    \end{aligned}
\end{equation}

We integrate first over all $t$ (note how we do not need $t\rightarrow\infty$ but $t\in(-\infty,\infty)$),

\begin{equation}
    \begin{aligned}
        f=\frac{2m}{\hbar}&\int dk'\; \tilde\psi(k')\delta(k'^2-k^2)\\
        \partial_xf=i\frac{2m}{\hbar}&\int dk'\; k'\tilde\psi(k')\delta(k'^2-k^2)
    \end{aligned}
\end{equation}

We expand the Dirac delta functions,

\begin{equation}
    \begin{aligned}
        f=\frac{m}{\hbar k} &\int dk'\; \tilde\psi(k')\left[\delta(k'+k)+\delta(k'-k)\right]\\
        \partial_xf=i\frac{m}{\hbar k} &\int dk'\; k'\tilde\psi(k')\left[\delta(k'+k)+\delta(k'-k)\right]
    \end{aligned}
\end{equation}

And integrate over $k'$,

\begin{equation}
    \begin{aligned}
        f=\frac{m}{\hbar k}&\left[\tilde\psi(k)+\tilde\psi(-k)\right]\\
        \partial_xf=i\frac{m}{\hbar}&\left[\tilde\psi(k)-\tilde\psi(-k)\right]
    \end{aligned}
\end{equation}

Plugging these results into Eq. \ref{eqcont} yields,

\begin{equation}
    \begin{aligned}
        \tilde\psi(k)=\frac{\hbar}{2m}\bigg(\frac{m}{\hbar}&\left[\tilde\psi(k)+\tilde\psi(-k)\right]\\
        +\frac{m}{\hbar}&\left[\tilde\psi(k)-\tilde\psi(-k)\right]\bigg)
    \end{aligned}
\end{equation}

which is expressly true. The only point of contention is when $k=0$ which represents a constant component of the wavefunction. In this case there is no flux moving through the VD and thus there is no propagation direction. One may either consider the zero point as unimportant or they may take it to simply be the average value of the wavefunction at the VD.

This proof's statement may be seen in another perspective. The sufficient conditions for a well-defined Schr\"odinger-like problem,

\begin{equation}
    i\partial_t\psi+a\partial_x^2\psi=0
\end{equation}
\begin{equation*}
    \psi:\mathbb{R}^2\rightarrow\mathbb{C},\;\;x,t\in(-\infty, \infty),\;\; a>0
\end{equation*}

are either the initial wavefunction over all space, $\psi(x,0)$, or the wavefunction and its derivative at one point for all time, $\psi(0,t)$, $\psi'(0,t)$. In fact this should be applicable for any dispersion relation which has two or zero real solutions $k_0$ for $\omega(k_0)=\omega_0$ for all $\omega_0\in\mathbb{R}$ except for some number of $k_0\in\mathbb{R}$ where $\frac{d\omega}{dk}(k_0)=0$ which cannot be resolved. This is the case for $k=0$ with the quadratic dispersion relation.

\section{Benchmarking}

We will benchmark the CVD, QVD, and BQVD against the exact initial distribution $|\tilde\psi(k)|^2$ numerically with a few examples. We use the time-dependent Schr\"odinger equation in atomic units ($\hbar=m=1$) with spacing $\Delta x=0.1$, $\Delta t=1$ and propagate using the operator-splitting Fourier method (periodic boundary conditions) from $t=0$ to $t=1000$. This method of propagation is exact for a constant potential.

\subsection{Single Gaussian Wavepacket}

We use a Gaussian wavepacket with initial conditions $x_0=-250$, $\sigma_x=50$, and $k_0=1$. The wavepacket passes through the measurement point, $x=0$, only once in the calculation's duration. For the BQVD the second evaluation point is at $x=\Delta x$ for all the proceeding cases. For the probability current the derivative is calculated using the points $x=\pm\Delta x$.

\begin{figure}[!htb]
    \centering
    \includegraphics[width=0.9\linewidth]{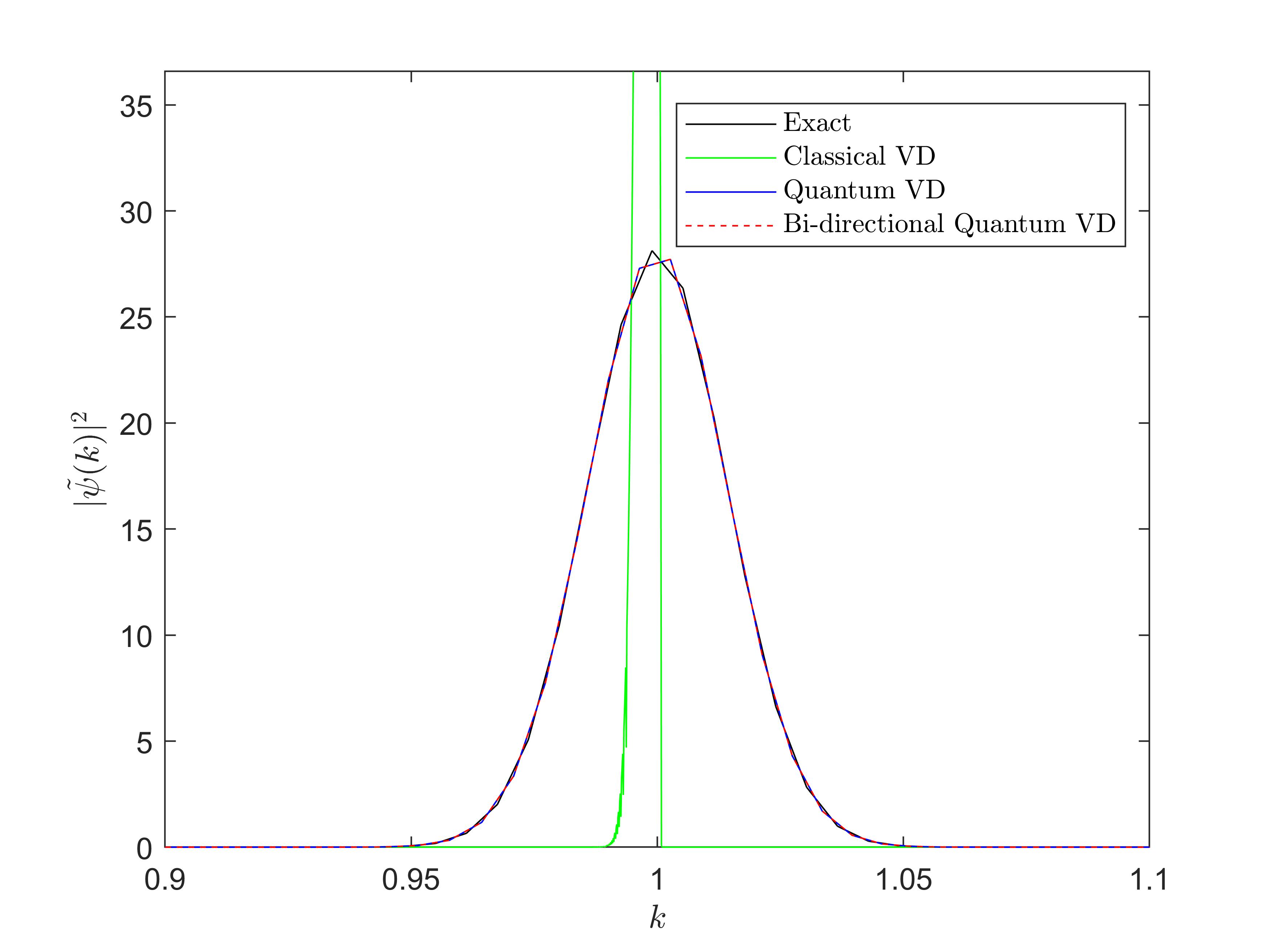}
    \caption{The exact (black), CVD (green), QVD (blue), and BQVD (red dashed) momentum spectra for a single Gaussian wavepacket. The QVD and BQVD overlap exactly and agree well with the exact result. The CVD measures primarily the central momentum as the wavepacket has not yet fully dispersed.}
    \label{fig:singgaus}
\end{figure}

The results of this calculation are shown in Figure \ref{fig:singgaus}. The QVD and BQVD results agree exactly and coincide well with the exact result. Provided infinitesimal $dt$ and $dx$ these three results should exactly match.

The CVD's observed momenta are confined near to the central momentum of the wavepacket. This is because the measured probability current divided by the probability density for a non-dispersive Gaussian wavepacket is constant everywhere, and thus the CVD only measures approximately a single value of momentum for all time.

For a dispersive wavepacket the momentum distribution is dependent on when it is measured -- using a CVD near to the wavepacket's starting point results in a sharper distribution, while measuring far away results in a broader distribution. This is because, as the wavepacket disperses, the front end will have higher momenta and the tail end will have lower momenta; with more travel time this effect of dispersion grows. As the travel time increases the wavepacket approaches a wave with slowly changing amplitude and phase gradient, ultimately resulting in a more accurate measurement. \cite{Feuerstein_2003}

\subsection{Two Counter-propagating Wavepackets}

\begin{figure}[!htb]
    \centering
    \includegraphics[width=0.9\linewidth]{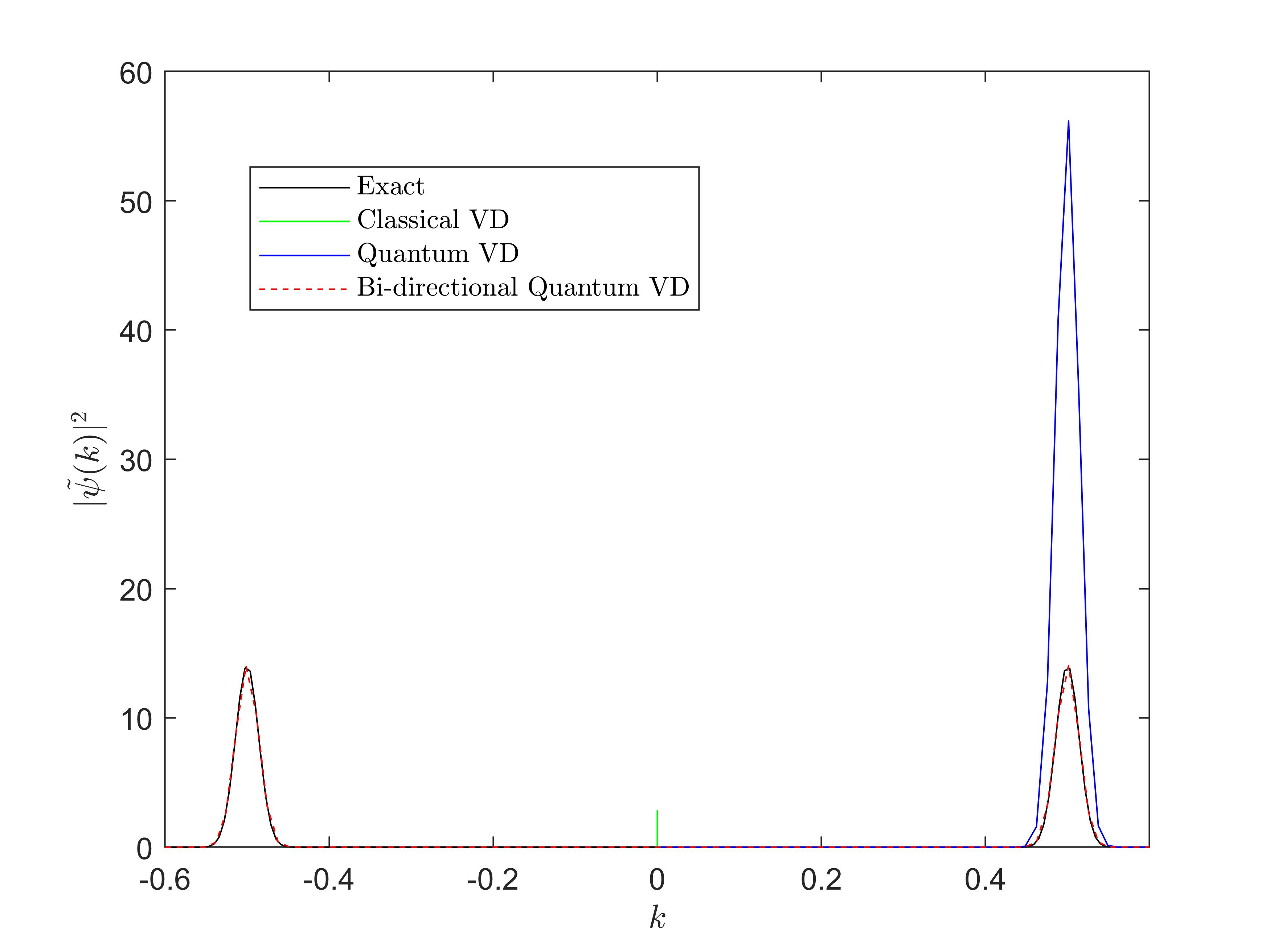}
    \caption{The exact (black), CVD (green), QVD (blue), and BQVD (red dashed) momentum spectra for two opposing wavepackets of symmetric phase. The BQVD agrees well with the exact result. The QVD, unable to discern between the directions of the two wavepackets, constructively interferes the two into the positive momentum value. The CVD measures zero probability current and therefore results in a small spike at $k=0$.}
    \label{fig:opgaus}
\end{figure}

Here we use two Gaussian wavepackets of opposite momentum $k_0=\pm0.5$, $x_0=\mp 250$, $\sigma_x=50$, and symmetric phase. The measurements are shown in Figure \ref{fig:opgaus}.

The BQVD agrees well with the exact result. As there is net zero probability current measured at the VD position the CVD measures zero momentum to numerical precision. The QVD cannot discern between rightward and leftward moving wavepackets so it combines the two at the same positive momentum, leading to a constructive interference in the spectrum. The QVD is also sensitive to the relative phase of the two wavepackets.

\begin{figure}[!htb]
    \centering
    \includegraphics[width=0.9\linewidth]{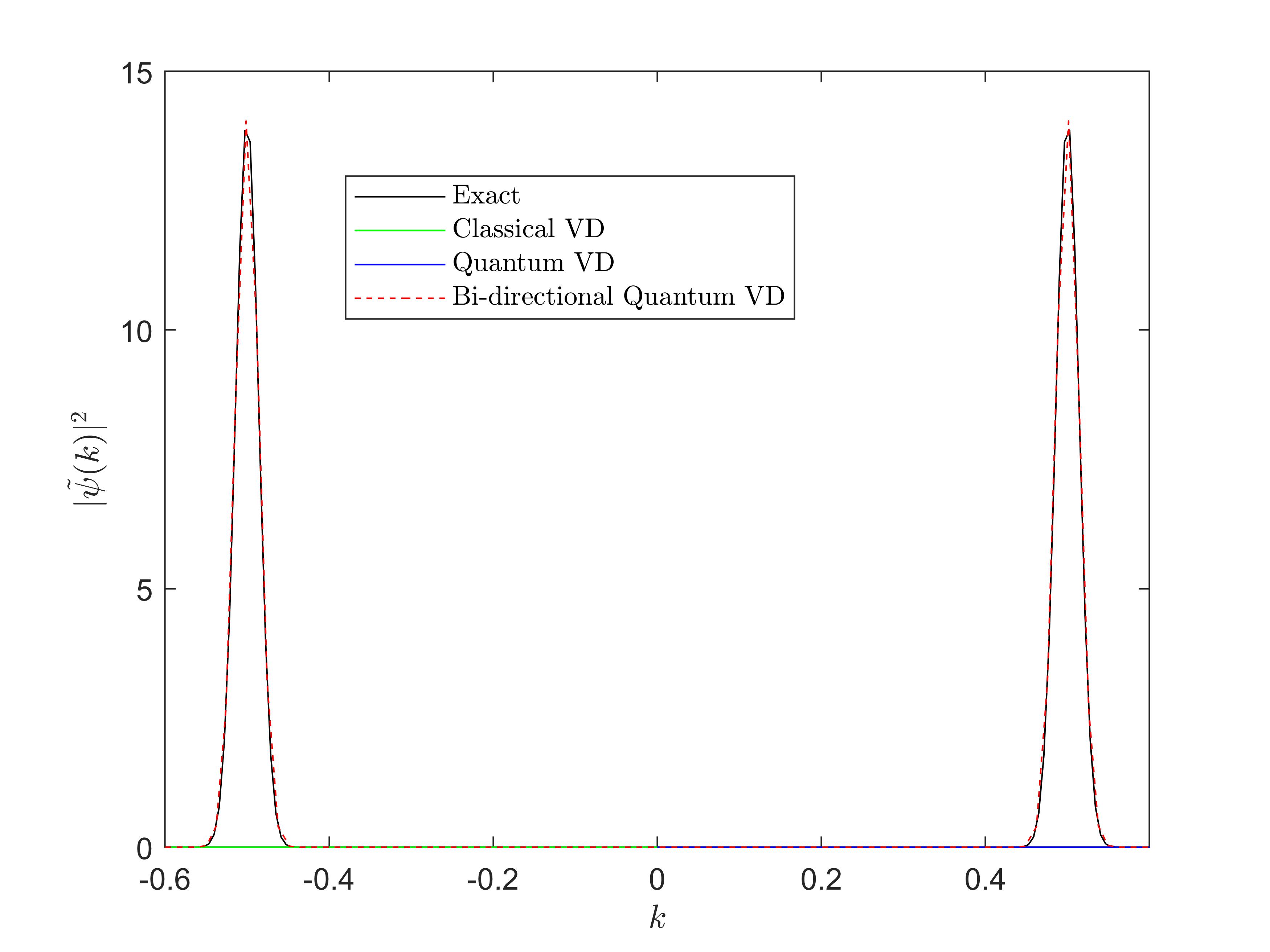}
    \caption{The exact (black), CVD (green), QVD (blue), and BQVD (red dashed) momentum spectra for two opposing wavepackets of anti-symmetric phase. The BQVD agrees well with the exact result. The CVD and QVD are near-zero everywhere due to the zero current observed at $x=0$ and due to the destructive interference of the two wavepackets, respectively.}
    \label{fig:opgausanti}
\end{figure}

 If we instead start with anti-symmetric states the wavefunction at $x=0$ is identically zero and so the QVD would measure no wavefunction, otherwise seen as a total destructive interference in the spectrum. As the BQVD takes the wavefunction's value at two points into consideration it is not limited by this special case, and in fact performs just as well as seen in Figure \ref{fig:opgausanti}. 

\subsection{Two Co-propagating Wavepackets}

\begin{figure}[!htb]
    \centering
    \includegraphics[width=0.9\linewidth]{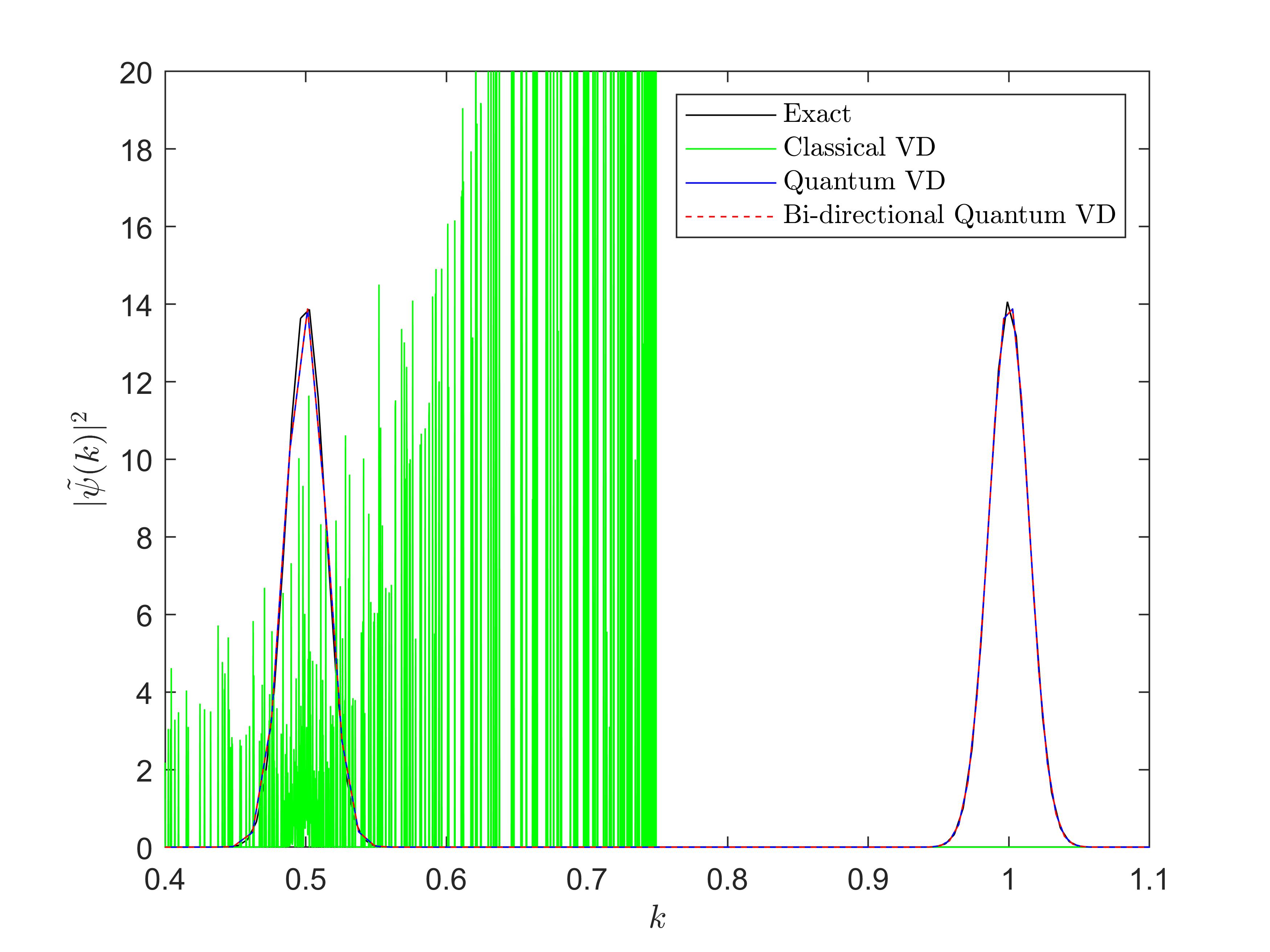}
    \caption{The exact (black), CVD (green), QVD (blue), and BQVD (red dashed) momentum spectra for two co-propagating wavepackets. The QVD and BQVD coincide well with the exact result. The CVD measures a slew of momenta as the two wavepackets rapidly interfere at the measurement point.}
    \label{fig:coprop}
\end{figure}

In this case we have two Gaussian wavepackets, one with central momentum $k_0=1$ and starting position $x_0=-250$ and the other with $k_0=0.5$ and $x_0=-125$. Their centroids then meet at $x=0$ and $t=250$. Both Gaussians use $\sigma_x=50$. These measurements are shown in Figure \ref{fig:coprop}. 

Both the QVD and BQVD coincide well with the exact result. The CVD measures a slew of momenta below and up to the average momentum $k=0.75$. Due to the interference of the two wavepackets the CVD is unable to resolve either wavepacket.

\section{Conclusion}

We have shown by proof and by example that the BQVD can reproduce the exact momentum distribution for a wavefunction propagating in a constant potential field. It avoids the interference problem of counter-propagating wavepackets found with the standard QVD and the incoherence problem with the CVD in all cases. It may potentially be extended to any Schr\"odinger-like equation that has a dispersion relation with two or zero real solutions while still requiring two measurement points. While the BQVD may be used generally as a virtual spectrometer, we note that the standard QVD performs just as well provided that propagation is restricted to one direction. This may be done by placing the VD at near the simulation edge where there is an absorptive boundary condition as is typically done for strong field emission measurements. Additionally the BQVD and QVD do not produce instantaneous spectra and therefore can not straightforwardly be used as a quantum-classical boundary. Finally, we have only proven equivalence in 1-D for the BQVD (and by extension $n$-D hyperplanar boundaries) whereas the CVD has been proven for $n$-D closed surfaces \cite{cvdproof}, given the momentum-spatial correlation requirement of $t\rightarrow\infty$.

\section{Acknowledgements}

This research was funded by the Center for Bright Beams, National Science Foundation Grant No. PHY-1549132.

\printbibliography

\end{document}